\documentclass[twocolumn]{aastex62}

\usepackage{longtable}
\usepackage{graphicx}
\usepackage{amsmath,amssymb}
\usepackage{color}
\usepackage{units}
\usepackage{epstopdf}
\usepackage{hyperref}
\usepackage{multirow}
\usepackage{float}
\usepackage{tikz}
\usetikzlibrary{shapes,arrows}

\usepackage{listings}
\usepackage{color}

\definecolor{dkgreen}{rgb}{0,0.6,0}
\definecolor{gray}{rgb}{0.5,0.5,0.5}
\definecolor{mauve}{rgb}{0.58,0,0.82}

\newcommand{\beq}{\begin{equation}}
\newcommand{\eeq}{\end{equation}}
\newcommand{\bdm}{\begin{displaymath}}
\newcommand{\edm}{\end{displaymath}}

\definecolor{Gray}{gray}{0.9}
\definecolor{orange}{rgb}{0.9,0.5,0}

\graphicspath{{./plots/}}
\begin{document}

\title{Optimizing searches for electromagnetic counterparts of gravitational wave triggers}

\author{Michael W. Coughlin}
\affil{Division of Physics, Math, and Astronomy, California Institute of Technology, Pasadena, CA 91125, USA}

\author{Duo Tao}
\affil{Carleton College, Northfield, MN 55057, USA}

\author{Man Leong Chan}
\affil{University of Glasgow, Glasgow G12 8QQ, United Kingdom}

\author{Deep Chatterjee}
\affil{University of Wisconsin-Milwaukee, Milwaukee, WI 53201, USA}

\author{Nelson Christensen}
\affil{Carleton College, Northfield, MN 55057, USA}
\affil{Artemis, Universit\'e C\^ote d'Azur, Observatoire C\^ote d'Azur, CNRS, CS 34229, F-06304 Nice Cedex 4, France}

\author{Shaon Ghosh}
\affil{University of Wisconsin-Milwaukee, Milwaukee, WI 53201, USA}

\author{Giuseppe Greco}
\affil{INFN, Sezione di Firenze, I-50019 Sesto Fiorentino, Firenze, Italy}
\affil{Universit a degli Studi di Urbino 'Carlo Bo,' I-61029 Urbino, Italy}

\author{Yiming Hu}
\affil{TianQin Research Center for Gravitational Physics, Sun Yat-sen University, Tangjiawan, Zhuhai 519082, Guangdong, P. R. China}

\author{Shasvath Kapadia}
\affil{University of Wisconsin-Milwaukee, Milwaukee, WI 53201, USA}

\author{Javed Rana}
\affil{Inter-University Centre for Astronomy and Astrophysics, Pune 411007, India}

\author{Om Sharan Salafia}
\affil{INAF - Osservatorio Astronomico di Brera Merate, via E. Bianchi 46, I–23807 Merate, Italy}

\author{Christopher Stubbs}
\affil{Department of Physics, Harvard University, Cambridge, MA 02138, USA\\
Department of Astronomy, Harvard University, Cambridge MA 02138, USA}

\begin{abstract}
With the detection of a binary neutron star system and its corresponding electromagnetic counterparts, a new window of transient astronomy has opened. 
Due to the size of the error regions, which can span hundreds to thousands of square degrees, there are significant benefits to optimizing tilings for these large sky areas.
The rich science promised by gravitational-wave astronomy has led to the proposal for a variety of tiling and time allocation schemes, and for the first time, we make a systematic comparison of some of these methods.
We find that differences of a factor of 2 or more in efficiency are possible, depending on the algorithm employed.
For this reason, for future surveys searching for electromagnetic counterparts, care should be taken when selecting tiling, time allocation, and scheduling algorithms to maximize the probability of counterpart detection.

\end{abstract}


\section{Introduction}
\label{sec:Intro}

The era of multi-messenger gravitational-wave astronomy has arrived with the detection of GW170817 \citep{AbEA2017b} by Advanced LIGO \citep{aLIGO} and Advanced Virgo \citep{adVirgo} coincident with the detection of both a short gamma-ray burst (SGRB) \citep{AbEA2017c,AbEA2017d,AbEA2017e} and kilonova detected in coincidence \citep{AbEA2017f,AbEA2017g}. 
This work is the culmination of significant effort expended in the search for the electromagnetic counterpart of the gravitational waves found by compact binary black hole systems \citep{AbEA2016a,AbEA2016g,AbEA2017}.

There has been significant optimism for the potential electromagnetic counterparts for emission from binary neutron star and black hole - neutron star systems across timescales and wavelengths \citep{Nakar2007,MeBe2012}. 
A  kilonova, arising from sub-relativistic ejecta, in particular has bolometric luminosities of $\approx 10^{40}-10^{42}$\,ergs/s \citep{MeBa2015,BaKa2013} (GW170817 peaked at $\approx 10^{42}$\,ergs/s \citep{SmCh2017})
and color and durations dependent on the physical conditions of the merger \citep{MeMa2010,KaBa2013,BaKa2013,TaHo2013,KaFe2015,BaKa2016,Me2017}.

The scientific output from a joint gravitational-wave and electromagnetic observation is significant, as the detection of a kilonova coincident with a gravitational wave allows for the exploration of the neutron star equation of state \citep{BaBa2013} and r-process nucleosynthesis in the unbound ejecta from a merger involving a neutron star \citep{MeBa2015,JuBa2015,RoLi2017,WuFe2016}. It also allows for a distance-ladder independent measurement of the Hubble Constant \cite{AbEA2017h}.
In addition, the joint observation with a short gamma-ray burst confirms these phenomena are driven by compact binary mergers, but also allow for the study of their beaming, energetics, and galactic environment \citep{MeBe2012}.

To facilitate the detection of gravitational-wave counterparts, probability skymaps as a function of sky direction and distance are released for gravitational wave triggers produced by the detectors \citep{SiPr2014,BeMa2015}. 
Due to the significant sky coverage required to observe the gravitational-wave sky localization regions, usually spanning $\approx 100\,\textrm{deg}^2$, techniques to optimize the followup efforts are of significant utility \citep{Fair2009,Fair2011,Grover:2013,WeCh2010,SiAy2014,SiPr2014,BeMa2015,EsVi2015,CoLi2015,KlVe2016}.
Given the large sky localization regions involved, wide-field survey telescopes have the best opportunities to make a detection. 
The Panoramic Survey Telescope and Rapid Response System (Pan-STARRS) \citep{MoKa2012}, Asteroid Terrestrial-impact Last Alert System (ATLAS) \citep{Ton2011}, the intermediate Palomar Transient Factory (PTF) \citep{RaSh2009} and (what will become) the Zwicky Transient Facility (ZTF), BlackGEM \citep{BlGr2015} and the Large Synoptic Survey Telescope (LSST) \citep{Ivezic2014} are all examples of such systems.
For example, Pan-STARRS has a 7$\textrm{deg}^2$ field of view (FOV), achieving a 5 $\sigma$ limit of 21.5 (AB mag) in the i band in a 45 second exposure. ATLAS has a 29.2$\textrm{deg}^2$ field of view, achieving a 5 $\sigma$ limit of 18.7 in the cyan band in a 30 second exposure. For comparison, LSST will have a $9.6\textrm{deg}^2$ FOV and will require a 21\,s r-band exposure length to reach 22 mag.

Due to the significant difference in telescope configurations, including FOV, filter, typical exposure times, and limiting magnitudes, in addition to placement on the earth and therefore different seeing and sky conditions, optimizing gravitational wave followups for generic telescopes is difficult. Therefore, in the following, we will take the telescopes mentioned above as examples. 

For this reason, we have created a codebase named \emph{gwemopt} (Gravitational Wave - Electromagnetic OPTimization) that utilizes methods from a variety of recent papers geared towards optimizing efforts of followup. We employ methods to read gravitational-wave skymaps and the associated information made available from GraceDB \footnote{https://gracedb.ligo.org}, in addition to information about the telescopes to tile the sky, allocate available telescope time to the chosen fields, and schedule that time in a way that optimizes based on expected lightcurves.
In section~\ref{sec:algorithm}, we describe the algorithm.
In section~\ref{sec:performance}, we describe the performance of the algorithms.
In section~\ref{sec:conclusions}, we offer concluding remarks and suggest directions for future research.

\section{Algorithm}
\label{sec:algorithm}

Figure~\ref{fig:flowchart} shows the flowchart for the \emph{gwemopt} pipeline, developed to optimize the efforts of electromagnetic followup of gravitational-wave events. \emph{gwemopt} is developed in python, which has the benefit of interfaces to both LIGO's gravitational-wave candidate event database (GraceDB) and HEALPIX \citep{GoHi2005}, the format in which LIGO reports skymaps.
In the following, we will show the command line syntaxes required to reproduce the results at the beginning of each section.

\emph{gwemopt} uses events provided by gracedb in addition to information about the telescopes for creating tiles and optimize time allocations in the fields.
It uses information about potential lightcurves from electromagnetic counterparts to schedule the available telescope time.
In the following, we will describe the calculations that go into creating tiling, time allocations, and observing sequences from the skymaps.
We will account for both diurnal and observational constraints and have the possibility of imaging over many nights.

\tikzstyle{decision} = [diamond, draw, fill=blue!20,
    text width=4.5em, text badly centered, node distance=3cm, inner sep=0pt]
\tikzstyle{block} = [rectangle, draw, fill=blue!20,
    text width=5em, text centered, rounded corners, minimum height=3em]
\tikzstyle{line} = [draw, -latex']
\tikzstyle{cloud} = [draw, ellipse,fill=red!20, node distance=3cm,
    minimum height=2em]
\tikzstyle{emptyblock} = [rectangle, minimum height=3em]

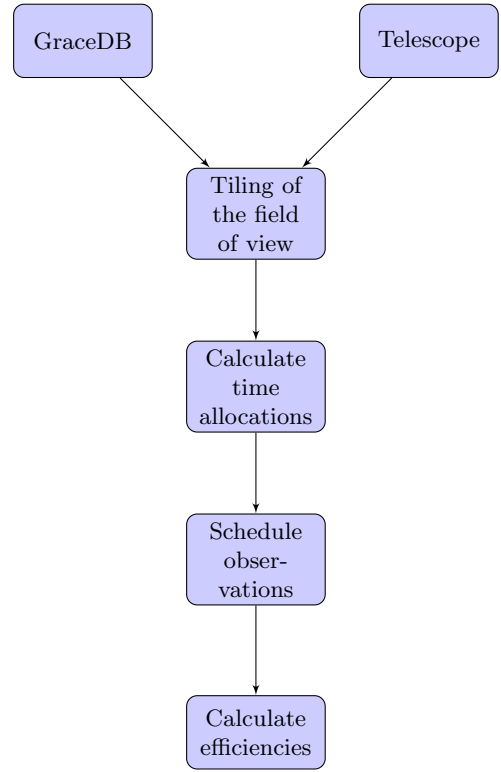
\begin{figure}[t]
 \begin{center}
 \begin{tikzpicture}[node distance = 2.3cm, auto]
    \node [emptyblock] (init) {};
    \node [block, left of=init] (GraceDB) {GraceDB};
    \node [block, right of=init] (Telescope) {Telescope};
    \node [block, below of=init] (Tiling) {Tiling of the field of view};
    \node [block, below of=Tiling] (Time) {Calculate time allocations};
    \node [block, below of=Time] (Schedule) {Schedule observations};
    \node [block, below of=Schedule] (Efficiency) {Calculate efficiencies};
    \path [line] (GraceDB) -- (Tiling);
    \path [line] (Telescope) -- (Tiling);
    \path [line] (Tiling) -- (Time);
    \path [line] (Time) -- (Schedule);
    \path [line] (Schedule) -- (Efficiency);
 \end{tikzpicture}
 \end{center}
 \caption{A flow chart of the \emph{gwemopt} pipeline.}
 \label{fig:flowchart}
\end{figure}

\subsection{GraceDB}

\begin{figure}[t]
\hspace*{-0.5cm}
\centering
\includegraphics[width=3in]{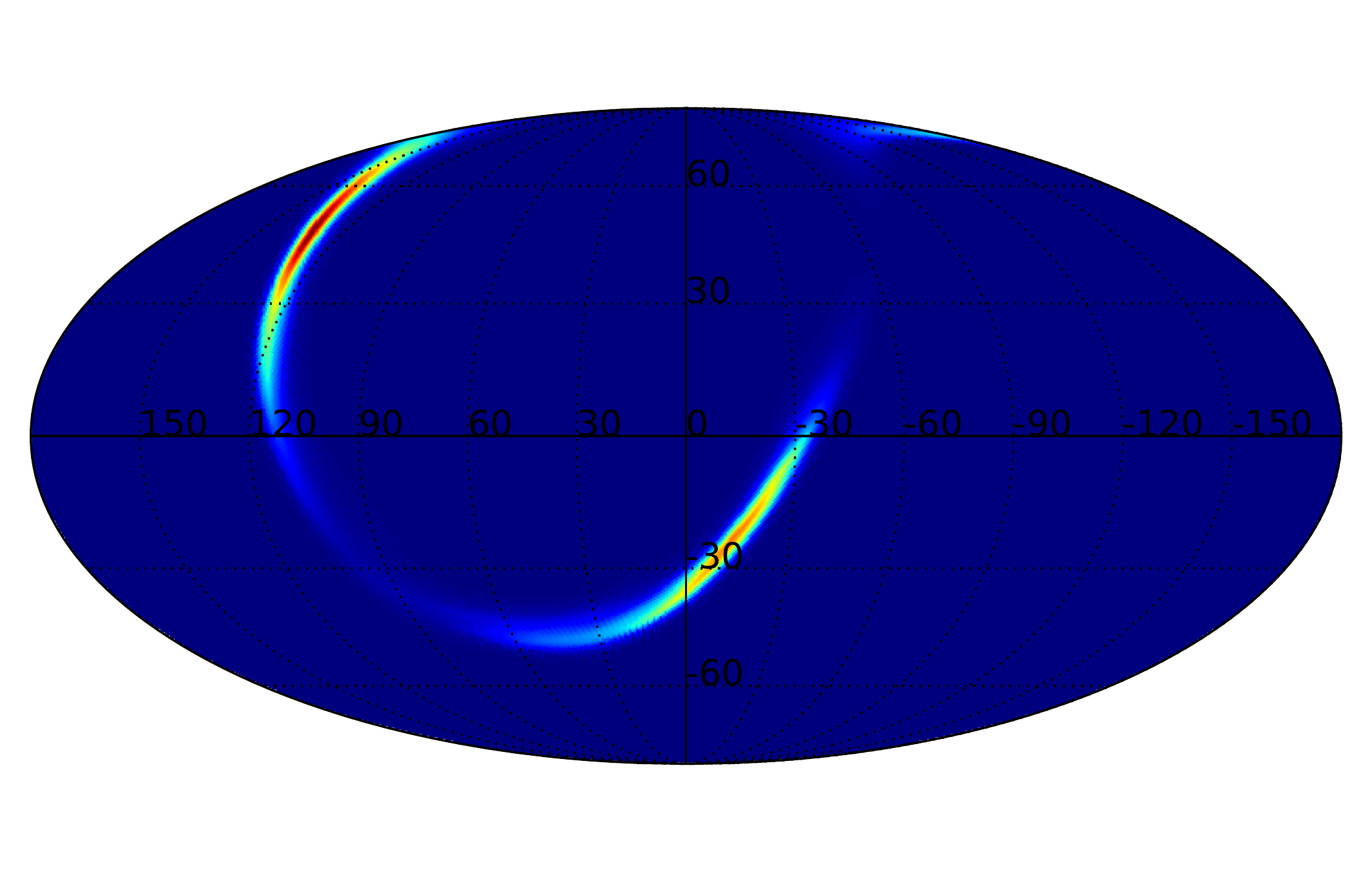}
\caption{The gravitational-wave likelihood $L_\textrm{GW}(\alpha,\delta,R)$ for GW170104.}
 \label{fig:skymap}
 \end{figure}

\begin{lstlisting}
python gwemopt_run --doEvent --do3D --event G268556
\end{lstlisting}

GraceDB is a service that provides information on candidate gravitational-wave events and the multi-messenger followups performed on them. An API is made available that allows for access to this information.
\emph{gwemopt} uses this API to access information pertinent for gravitational-wave followups.
First of all, it downloads the gravitational-wave skymap for a given event; an example is shown in Figure~\ref{fig:skymap}.
In addition, information such as the time of the event, the time delay between the time-of-arrival at the detectors, and EM bright information is noted.
\subsection{Telescope configuration}
\begin{lstlisting}
python gwemopt_run --doEvent --do3D --telescope LSST
\end{lstlisting}
\begin{table*}[t]
\scriptsize
\centering
\begin{tabular}{|c|c|c|c|c|c|c|c|c|}
\hline
Telescope & Latitude {[}deg{]} & Longitude {[}deg{]} & Elevation {[}m{]} & FOV {[}deg{]} & FOV shape & Filter & Exp. time {[}s{]} & Lim. Mag. \\ \hline
ATLAS          & 20.7204            & -156.1552           & 3055.0            & 5.46                                      & Square              & c      & 30.0                  & 18.7               \\ \hline
Pan-STARRS     & 20.7204            & -156.1552           & 3055.0            & 1.4                                      & Circle              & i      & 45.0                  & 21.5               \\ \hline
BlackGEM       & -29.2612           & -70.7313            & 2400.0            & 2.85 & Square              & g      & 300.0                 & 23.0               \\ \hline
LSST           & -30.1716           & -70.8009            & 2207.0            & 1.75                                      & Circle              & r      & 30.0                  & 24.4               \\ \hline
ZTF            & 33.3563            & -116.8648           & 1742.0            & 6.86  & Square              & r      & 30.0                  & 20.4               \\ \hline
\end{tabular}
\caption{Configuration of telescopes.}
\label{table:config}
\end{table*}
We require standardized configuration files for the telescopes to be analyzed. 
The information includes the filter being used, the limiting magnitude of the instrument and the exposure time required to achieve that magnitude, site location information, and information about the field of view shape and size. For the field of view, two options, square and circle are available, with the FOV being specified by the length of the square side and the radius of the circle. In addition, a tesselationFile is requested. This is especially useful for telescopes such as ZTF which use fixed telescope pointings which ensures the availability of reference images. In case a tesselation file is not available, one is automatically generated, described in the next section.
Configuration files for ATLAS, BlackGEM, LSST, PS1, and ZTF are available.
Table~\ref{table:config} provides the information assumed for these telescopes.\\
\subsection{Skymap tiling}
\begin{lstlisting}
python gwemopt_run --doEvent --do3D --doTiles --doPlots --tilesType ranked
\end{lstlisting}
There are a variety of algorithms in the literature for sky-map tiling, and the ones implemented in \emph{gwemopt} will be detailed below. The idea is to cover the sky with tiles the size of the telescope's field-of-view with minimal overlap. In some cases, these tiles are pre-determined by survey constraints in order to simplify difference imaging. In other cases, it is possible to optimize the tile locations based on the gravitational-wave skymaps, such that the tiles maximize the probability contained. In the following, we will check the difference between these tile locations to determine their effect. Due to the fields-of-view for these telescopes being in general much smaller than the probability region, the effect is expected to be relatively minimal.

\begin{figure*}
    \centering
    \includegraphics[width=3in]{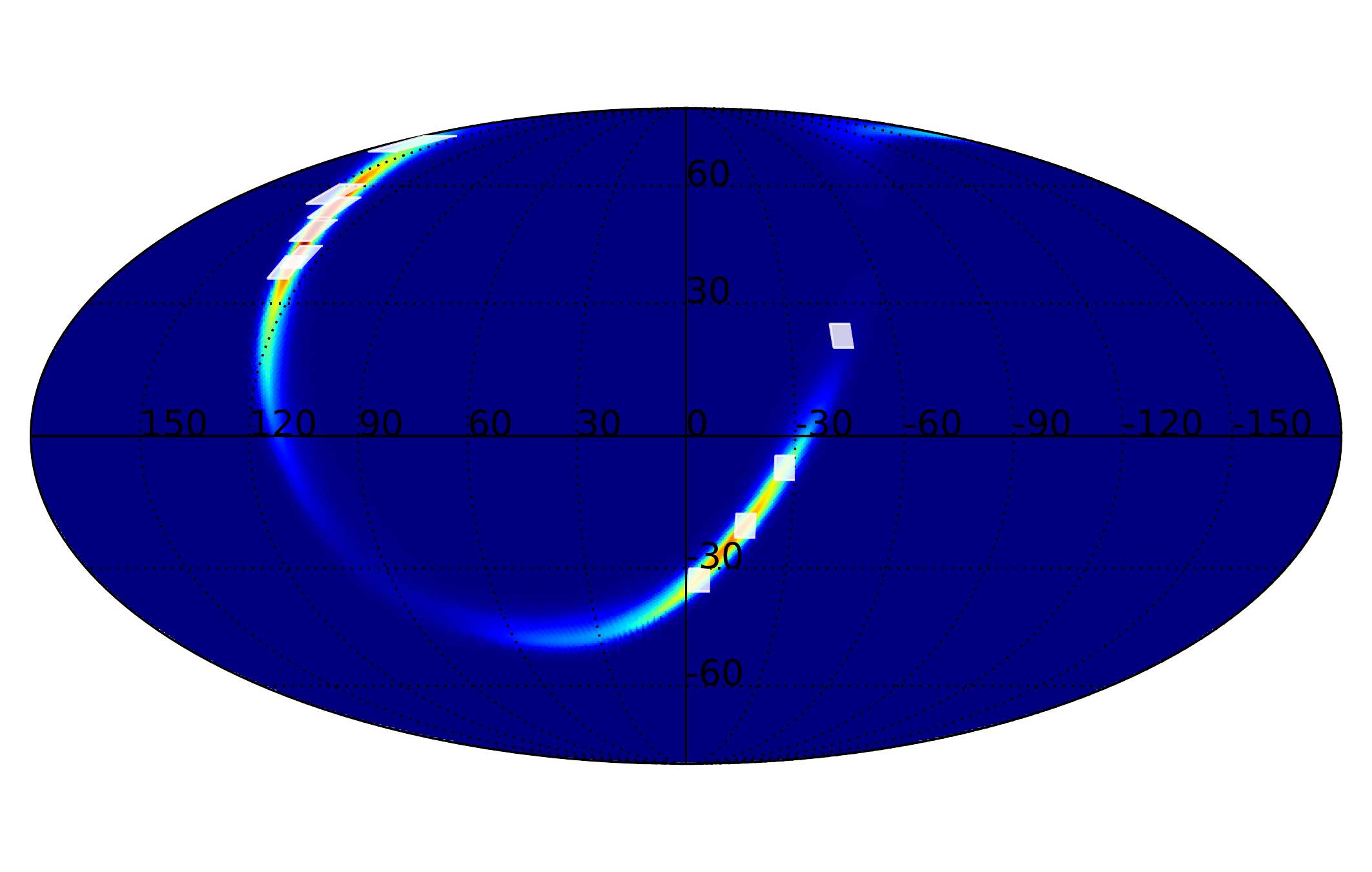}
    \includegraphics[width=3in]{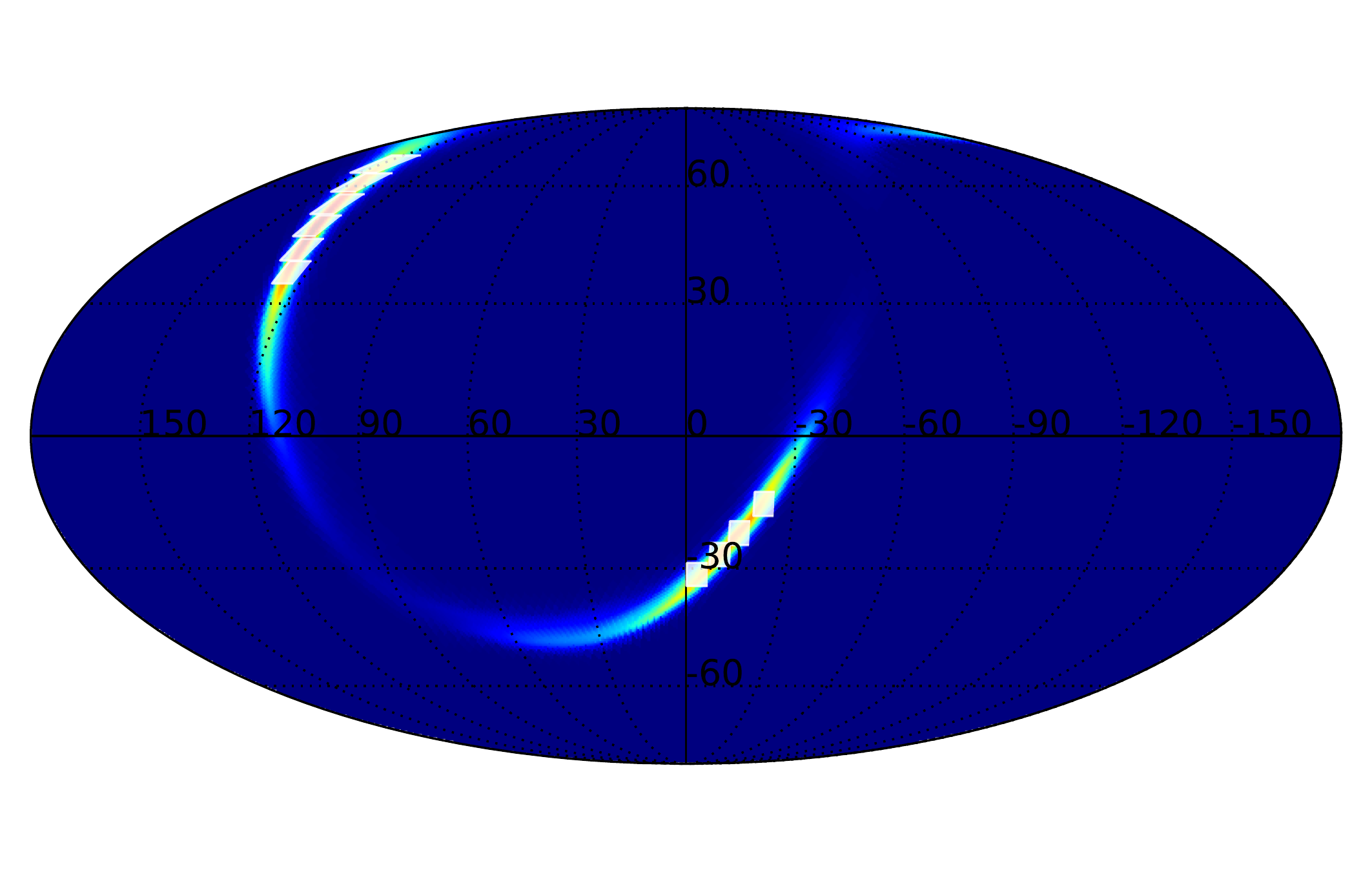} \\
    Greedy \hspace{2.6in}Hierarchical\\
    \includegraphics[width=3in]{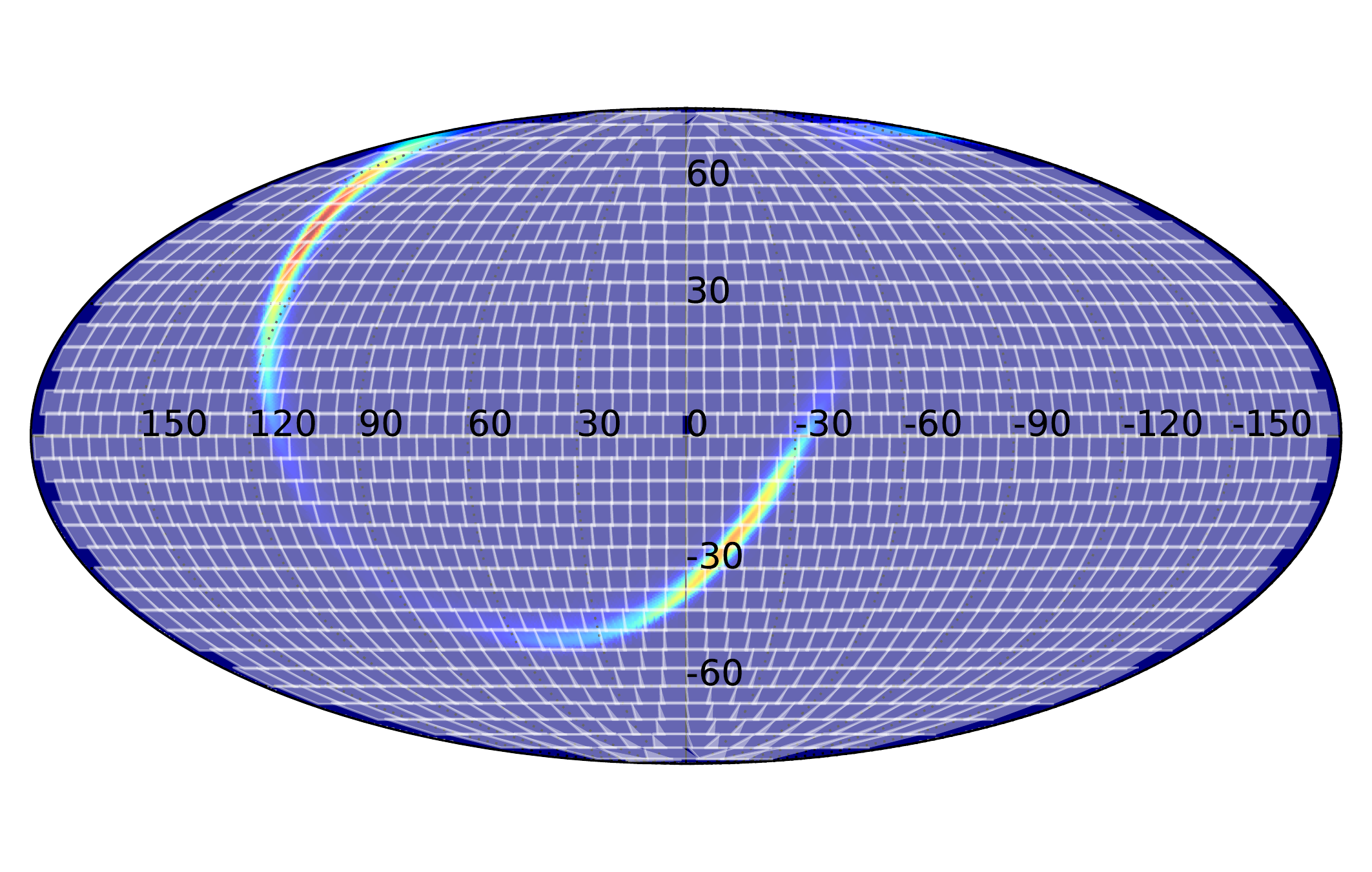}
    \includegraphics[width=3in]{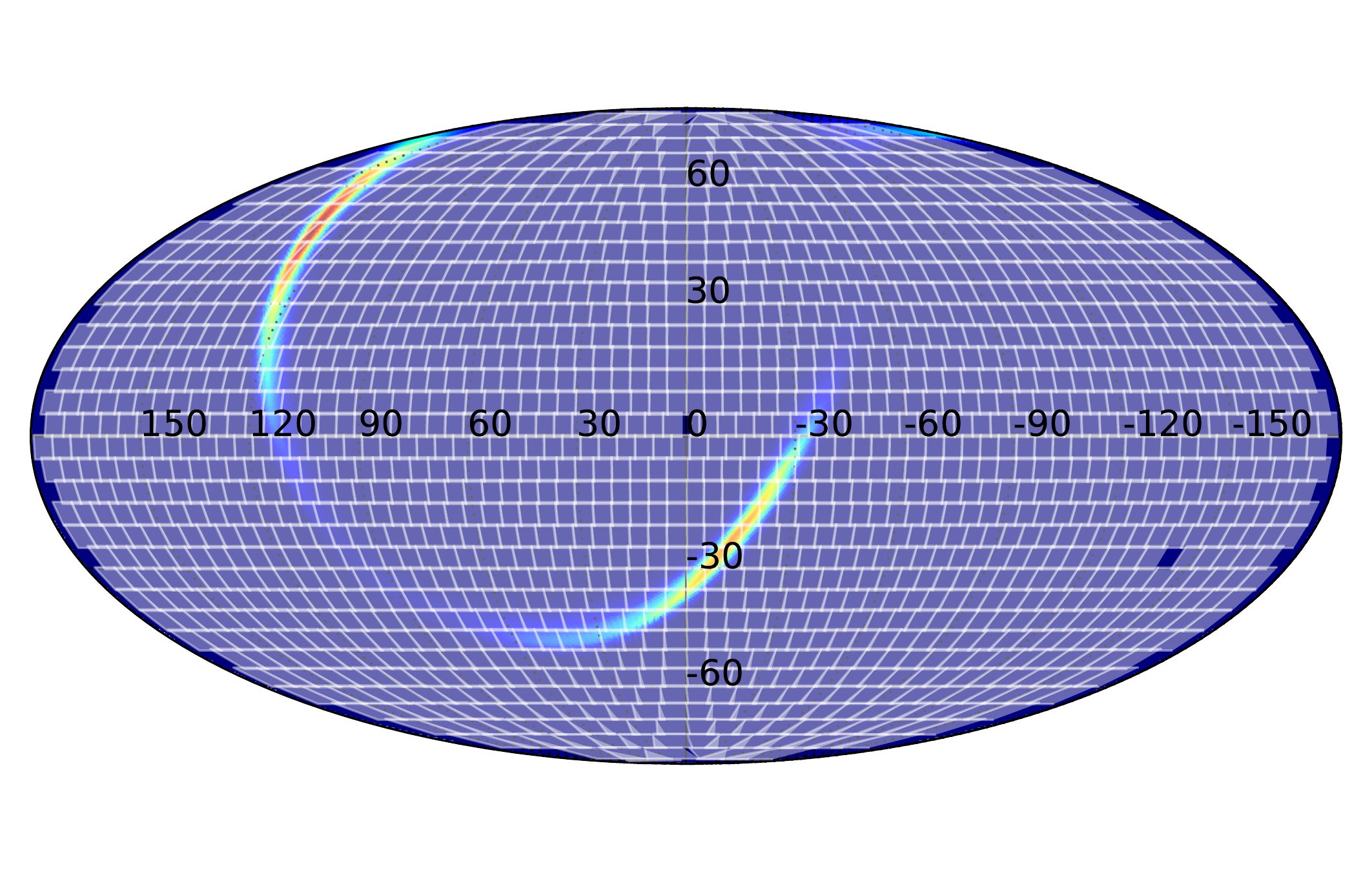} \\
    MOC \hspace{2.6in}Ranked
    \caption{Example outputs of different tiling algorithms. On the top left is the greedy version with $N_{\rm tiles}$ = 10, where $N_{\rm tiles}$ is the number of tiles employed, on the top right is the hierarchical version with the same, on the bottom left is the MOC skymap, and on the bottom right is the ranked skymap tiling.}
    \label{fig:tiling}
\end{figure*}

Gravitational-wave skymaps in general contain metrics that report the spatial probability of a gravitational-wave source lying within a certain location.
They are composed of HEALPIX arrays that encode either the 2D probability, in right ascension and declination, or 3D probability, which includes probability distributions for the distance.
They are reported in particular number of pixels, usually $\textrm{N}_\textrm{side} = 512$. 
This can introduce quantization errors, especially for small field-of-view telescopes. 
The --nside flag allows for the over- or under-sampling of the skymaps in the analysis.

There are four options related to skymap tiling currently available and defined below: MOC, ranked, hierarchical and greedy.
In the following, we will summarize the key features of each implementation, and referring the reader to the literature for further details.
The goal is to place each algorithm in the same mathematical formalism for straightforward comparisons.

\emph{MOC}. Multi-order coverage of healpix maps hierarchically predefines cells in order to specify arbitrary sky regions \citep{FeBo2014}. MOC is proposed in order to provide fast set operations between regions on the sky. In MOC, the spherical sky is recursively divided into 4 regions and each region is a diamond. 
The sphere is divided recursively into four diamonds. 
The division stops according to the resolution necessary for a particular usage.

Here are two relevant implementation details about MOC.
\begin{itemize}
\item MOC uses an equatorial coordinate system. 
\item MOC divides the sphere recursively into four diamonds. 
\item MOC indexes each tile as follows: the initial tile is numbered 0 on level 0. Then, when divided, we get tile indices of 0, 1, 2, 3 on level 1. More generally, if we start from a tile numbered M, its children will be numbered $M\times 4$, $M\times 4 + 1$, $M\times 4 + 2$,$M\times 4 + 3$ on the next level.
\end{itemize}

\emph{ranked}. \cite{GhBl2016} use pre-defined sky cells. This tiling scheme is also based on a grid system with grids of equal sizes such as the one used by MOC. The sizes of the grids are the same as the size of the telescope FOV. For each tile in the grid at $(\alpha_i, \delta_i)$, we calculate a double integral that accumulates the probability distribution in this tile, shown in Eq.\ref{eqn:tile_ranked},
\begin{equation}\label{eqn:tile_ranked}
T_{ij} = \int_{\alpha_i}^{\alpha_i+\Delta \alpha}\int_{\delta_i}^{\delta_i+\Delta \delta}L_\textrm{GW}(\alpha,\delta)d\Omega
\end{equation}
where $L_\textrm{GW}(\alpha,\delta)$ is the gravitational-wave likelihood.
Then, we rank all the tiles with their $T_{ij}$ and select from the top of the rankings until we reach the target probability of 95\%.

\emph{hierarchical}. A Multinest-based \citep{FeHo2009,FeGa2009,BuGe2014} optimization which optimizes tiles for a given skymap by placing them sequentially. This method starts by selecting the tile that covers the most probability. Then, it sets the probability in that tile to be zero before going to the next iteration, when it again selects the tile that covers the most probability. It stops until a user-specified number of tiles, $N_{\rm tiles}$, are selected. The tiles selected might overlap on the corners when there are higher probability distributions around that corner. 
 
\emph{greedy}. An emcee \citep{FoHo2013} based algorithm which optimizes tiles for a given skymap by placing them simultaneously. This method selects tiles that cover the highest probability altogether from the skymap. It ranks all possible tiles and selects from the top. Thus, the tiles selected by greedy method might overlap significantly when the probability distribution is concentrated.

\subsection{Time allocations}
\begin{lstlisting}
python gwemopt_run --doEvent --doPlots --doTiles --doSchedule --timeallocationType coverage
\end{lstlisting}
Once the tile locations have been assigned, whether dynamically or having been fixed previously, the next task is to assign time allocations to each tile, based on a variety of metrics. 
Because telescopes have fields of view that are in general significantly smaller than the probability region and typical exposure times for these telescopes are of order minutes (see Table~\ref{table:config}), it is not possible to image the entire probability region to interesting limiting magnitudes in a reasonable amount of time.
There are further constraints that arise from the diurnal cycle, observing time available for followup, limitations on the pointing a particular telescope is capable of, and the rise and set of tiles.
The following algorithms use a variety of methods to optimize the probability of imaging a counterpart.

The amount of time allocated is defined with a few constraints. 
First of all, time segments are generated based on the observing time allocated after the gravitational-wave event. 
In the following analyses, the default will be to assume the following 3 days are available.
The segments are then intersected with night time at the site of the particular telescope.
This defines the available segments. 
The time available for the analysis is then determined from these segments.
This assumes implicitly that the electromagnetic counterpart has not faded beyond detection limits in the time available. Some of the time allocation algorithms below will use models to determine when the counterpart is expected to be too faint to be detectable.

\begin{figure*}
    \centering
    \includegraphics[width=3in]{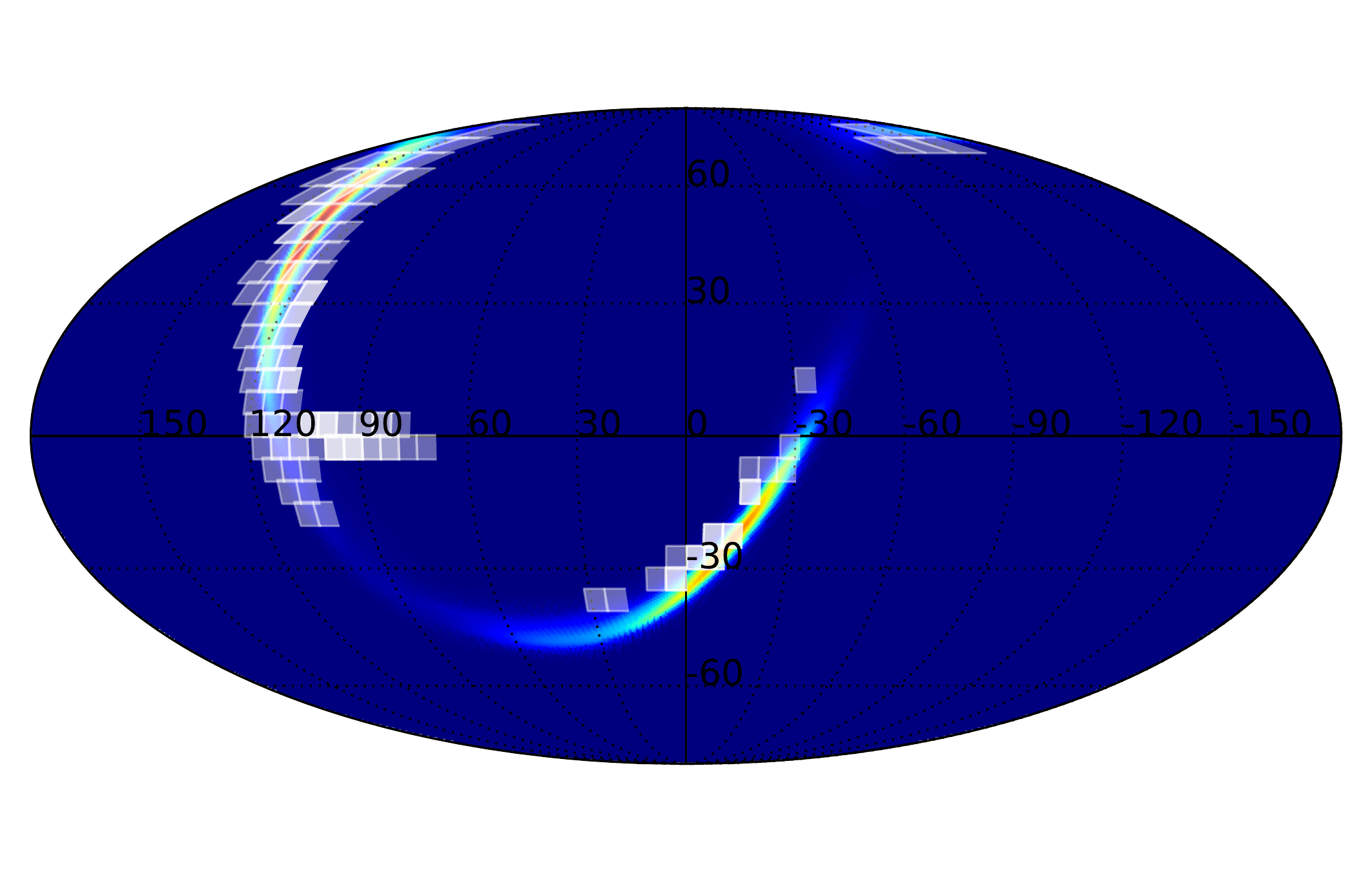}
    \includegraphics[width=3in]{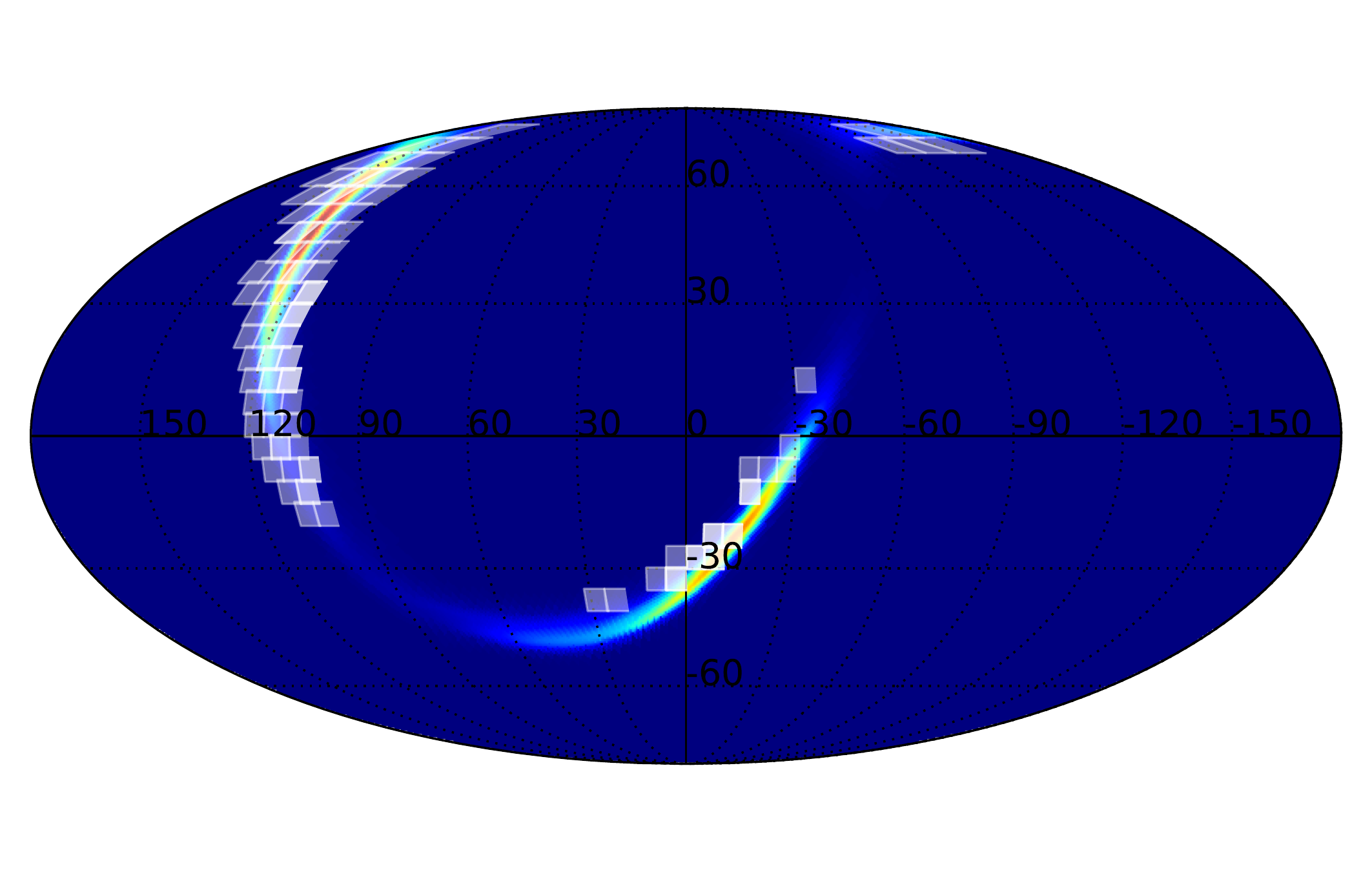}\\
    PEM\hspace{2.6in}Powerlaw\\
    \includegraphics[width=3in]{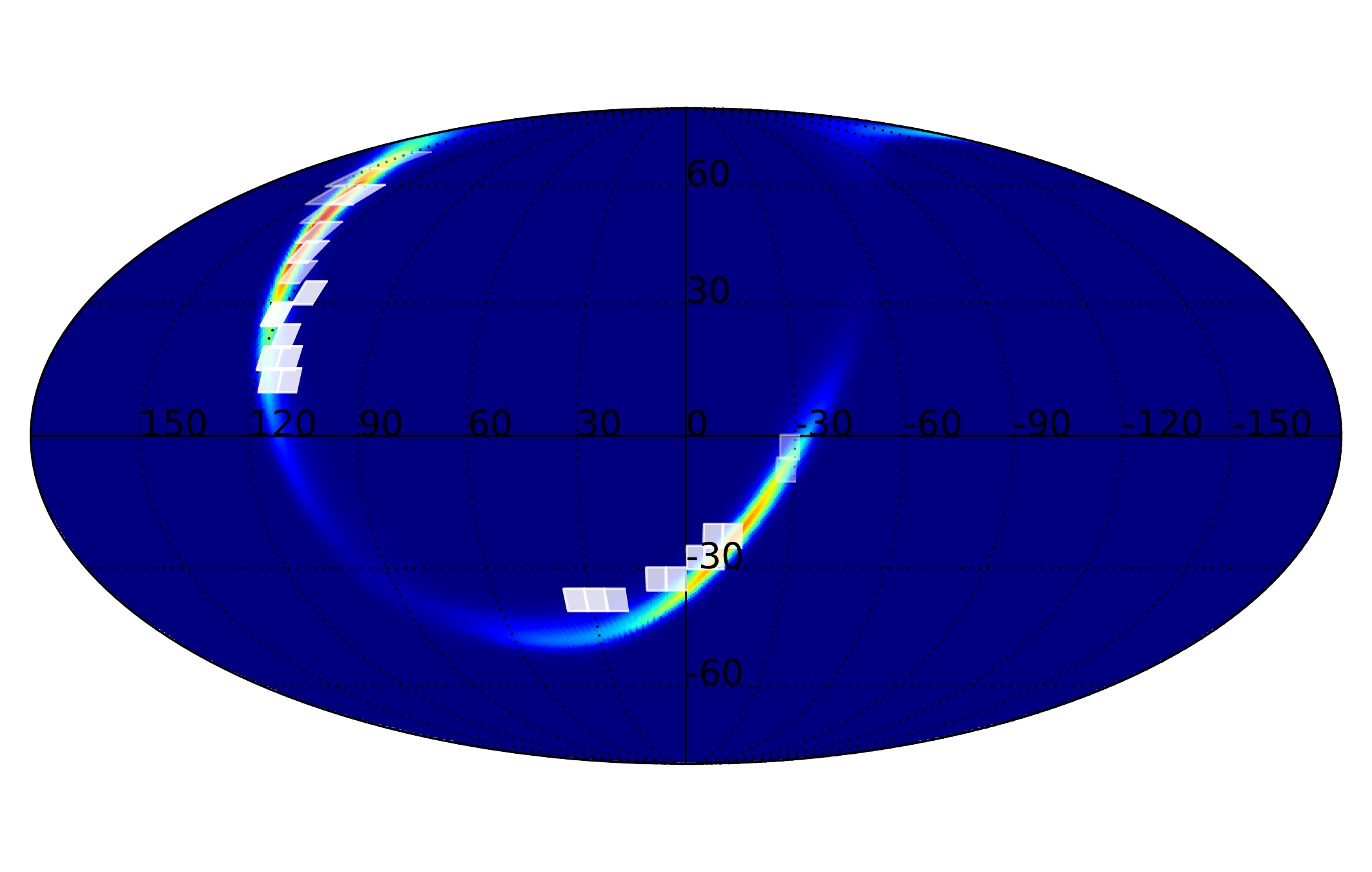}\\
    WAW
    \caption{Example outputs of different time allocation algorithms. On the top left is the tiles coverage with the PEM algorithm. On the top right is the tiles coverage with the Powerlaw algorithm. On the bottom is the tiles coverage with the WAW algorithm. In generating all of the plots, MOC algorithm is used.}
    \label{fig:timealloc}
\end{figure*}

There are four options related to time allocations as a function of sky location available, powerlaw, WAW (Where and when), PEM (Probability of electomagnetic counterpart), and coverage. Figure~\ref{fig:timealloc} shows examples of the powerlaw, WAW, and PEM types.

\emph{Coverage.} This is an option whereby coverage from existing surveys, including the right ascension and declination of the pointing and the limiting magnitude, are used.

\emph{Powerlaw.} Many searches have used a variation on simply scaling the time allocation proportional to the probability skymaps, a technique employed in the Powerlaw method below. \cite{CoSt2016a} derived scaling relations for the time allocated to any given field, $t_i$, given the graviational-wave likelihood. We showed that under certain assumptions, $t_i \propto \left(\frac{L_\textrm{GW}(\alpha_i,\delta_i)}{a(\alpha_i,\delta_i)}\right)^{2/3}$, where $L_\textrm{GW}(\alpha_i,\delta_i)$ is the gravitational-wave likelihood and $a(\alpha_i,\delta_i)$ is Galactic extinction.  While the powerlaw based analysis is straightforward, it does not account for the fact that the telescope must be sensitive enough to detect the counterpart. In this sense, this algorithm is the least model dependent. Although the detectability is model-dependent, both in the distances returned by the gravitational-wave detectors and the absolute magnitude of the sources, the following algorithms account for this in multiple ways.

The \emph{Powerlaw} algorithm optimizes the probability of detecting the transient with N observations, which is simply the sum of the probability of each observation. The expression is shown in Eq.\ref{eqn:powerlaw}:
\begin{equation}\label{eqn:powerlaw}
p_{tot}=\Sigma_{i=1}^N \frac{M_i}{M_{tot}}\frac{L_{GW}(\alpha_i, \delta_i, R_i)}{L_{GW_tot}}\frac{F_i(t_i)}{a(\alpha_i, \delta_i)},
\end{equation}
where $M_i$ is the mass for galaxy $i$; $M_{tot}$ is the total mass of galaxies in the field; $L_{GW}(\alpha_i, \delta_i, R_i)$ is the likelihood of the gravitational wave source in this galaxy; $F(t)$ is the luminosity as a function of allocated time; $a(\alpha_i, \delta_i)$ is the attenuation. In the following, we will simply scale the gravitational-wave likelihood and not the mass in the field. It is possible that this approximation could be improved using galaxy catalogs, although this introduces concerns about galaxy catalog completeness. Eq.\ref{eqn:powerlaw} is optimized with the constraint that the total observation time is limited, shown as
\begin{equation}
\Sigma_{i=1}^N t_i=T,
\end{equation}
where $T$ is the total observation time.

\emph{WAW} \cite{SoCo2017} use counterpart lightcurve models in the optical, infrared and radio constructed from information from the gravitational-wave signals to create a time- and sky location dependent probability for detecting electromagnetic transients. The WAW approach introduces time into the model by defining a concept of detectability. Detectability is the probability of detecting a light flux greater than the flux limit at position $\alpha$ and time $t$. Thus, by having detectability introduced, the algorithm can optimize "where" and "when" to schedule the observation based on $\alpha$ and $t$ with a greedy approach. The procedures of the algorithm are shown below.
\begin{enumerate}
\item The tiles are generated covering the confidence region based on the probability distribution, which comes from the gravitational wave signal.
\item The algorithm takes in the information encoded in the gravitational waves and computes the lightcurve $F_i(t)$ for each tile. 
\item Then, it computes the detectability as
\begin{equation}\label{eqn:detectability}
P(F(t) > F_{lim}|\alpha, S)\approx\Sigma_{i=1}^N\omega_i H(F_i(t)-F_{lim})
\end{equation}
where $H$ is the Heaviside function so if $F_i(t)$ is greater than $F_{lim}$, it is 1; otherwise it is 0. $F_i(t)$ is the light flux for position sample $i$ at time t. $F_{lim}$ is the limit flux. $\omega_i$ is the "inverse distance weight" that gives the contribution of the sample $i$ to position $\alpha$. The further away sample $i$ is from $\alpha$, the less it contributes. $\omega_i$ is normalized so that $\Sigma_{i=1}^N\omega_i=1$. 
\item For each tile, we find a time interval $[t_{E,\lambda},t_{L,\lambda}]$ when the detectability is greater than a threshold $\lambda$.
\item We start from the tiles that cover the most probability and arrange their observation times $[t_{E,\lambda},t_{L,\lambda}]$ if the time is available.
\end{enumerate}
This method optimizes the search by introducing detectability, defined as Eq.\ref{eqn:detectability} over the the three dimensional observation volume of direction and time, with the constraint that only one location can be observed at the same time.

\emph{PEM.} \cite{ChHu2017} optimize the number of fields to observe and their time allocations by adopting priors on the intrinsic luminosity of the sources and using knowledge of distance to the counterparts provided by the low-latency gravitational-wave searches through BAYESTAR \citep{SiPr2016} or high-latency parameter estimation from LALInference \citep{VeRa2015} for compact binary coalescence. More concretely, its input is the sky localization map and information about the telescope. It selects the tiles to observe with a greedy algorithm and allocates the observation time for each tile to maximizes the probability of detecting the EM counterpart of the GW event. The outputs are the tiles to be observed and the time allocated on each tile.

The procedures are shown below.
\begin{enumerate}
\item Based on the sky localization map, locate the tiles that cover the region enclosed by the contour of the target confidence level.
\item These N tiles are ranked based on the total probability covered.
\item We optimize the number of tiles selected and then the time allocation for each selected tile. For all k from 1 to n, we do the following:
\begin{enumerate}
\item the top k tiles from the rankings are selected. 

\item Eq.\ref{eqn:pem_obj} is optimized with Lagrange multiplier with the constraint of Eq.\ref{eqn:pem_const}. 
\begin{equation}\label{eqn:pem_obj}
P(D_{EM}|k)=\Sigma_{i=1}^{k\leq n}P(D_{EM}|\omega_i^{(k)}, \tau_i^{(i)},I)
\end{equation}
\begin{equation}\label{eqn:pem_const}
kT_0+\Sigma_{i=1}^k\tau_i^{(i)}=T
\end{equation}
Eq.\ref{eqn:pem_obj} is the total detection probability of all the tiles, and Eq.\ref{eqn:pem_const} is the constraint given by the observation resource of the telescope. 

Eq.\ref{eqn:pem_obj} is the sum of its detection probability of each tile: $\omega_i$ is the probability density; $\tau_i$ is the time allocated and $I$ is the parameters of the telescope. Its calculation is given in Eq.4 and 5 of \cite{ChHu2017}.

In Eq.\ref{eqn:pem_const}, $T_0$ is the time to adjust the telescope before each observation; $\tau_i$ is the time allocated to each tile; T is total observation resource.
\item We keep track of the best k, the tiles and the time allocation.
\end{enumerate}

\item The optimal tiles $\{\omega_i\}$ and their allocated times $\{\tau_i\}$ for $i=1...k$ are the output.
\end{enumerate}

\subsection{Scheduling}
\begin{lstlisting}
python gwemopt_run --doEvent --doPlots --doTiles --doSchedule --scheduleType weighted
\end{lstlisting}
Once the time allocated to each tile has been set, the next task is to schedule the observations that both best represent the time requested and optimize the times that are chosen in some way, for example, such that tiles are re-imaged at an approximately fixed cadence so as to measure possible lightcurve evolution or to go as deep as possible in one set.
Other optimizations might employ ordering based on airmass, as sources imaged through higher airmass will have lower signal-to-noise ratios.

There are three options related to scheduling observations, greedy, sear, and weighted.
The time that each tile is available for observation above the altitude limit is computed.
Using the set of segments available to the telescope, these tile-specific segments  are intersected with these segments to form a set of visibility segments for each tile.
This has the benefit of avoiding issues related to simply tracking the rise and set times of each tile.
To account for lunar sky brightness, we use a model from \cite{CoSt2016b}. 
Any tile whose sky brightness is increased by at least 1\,mag is excluded.

\emph{Greedy.} The simplest version of scheduling employs a schedule simply on the basis of probability contained. The idea is that higher ranked tiles are observed before lower ranked tiles based on this ranking scheme. \cite{RaSi2017} implemented a greedy algorithm whereby the field with the highest probability region in a given time window is observed. As this analysis did not include the possibility of multiple exposures for each pointing, it is modified in the analysis to include multiple exposures. The algorithm is as follows:

\begin{enumerate}
\item Construct a list of the tiles and number of exposures for each tile based on the time allocation algorithm utilized.
\item For each window, find the sky tiles that are in the current window: $T_0 + (j-1) T_{exp}$ and $T_0 + j T_{exp}$
\item Allocate the window to the sky tile with the greatest probability, and increment the number of exposures for that tile down by 1.
\end{enumerate}

\emph{sear (Setting Array).} The greedy algorithm has the short-coming that it does not account for site visibility. This motivates re-ordering the sequence such that as many tiles can be imaged as possible.
\cite{RaSi2017} also implemented a version whereby the rising and setting of tiles were accounted for. It uses the idea that observes high probability tiles first, subject to the condition that each tile from the observing sequence must be observed before it sets. The concept of imaging windows are used in this algorithm. We call a tile belongs to a window when the end of its observation time is in the window and denote the $i$th window with $W_i$. The algorithm uses the recursive relation between the optimal observation arrangement between the first $k$ windows $S_k$ and the first $k + 1$ windows $S_{k+1}$. The details are shown below. 
\begin{enumerate}
\item Consider the first window $W_1$ and initialize $S_1$ to be the tile that has the highest probability for $W_1$.
\item Move on to $W_2$ and find the two tiles $c_1$ and $c_2$ that have the greatest probability density.
\item Compare $c_1$ and $c_2$ with $S_1$ and act depending on the following conditions:
\begin{itemize}
\item If both $c_1$ and $c_2$ contain greater probability than $S_1$, set $S_2$ to be \{$c_1$, $c_2$\}.
\item Otherwise, put the tile with higher probability coverage between $c_1$ and $c_2$ into $S_2$, which becomes $S_3$.
\end{itemize}
We can see that either way, $S_2$ will have two elements.
\item Move on to the next observation windows until the last one. The only difference for the coming iterations from the descriptions above is that $S_k$ will have k elements.
\item Return the last set $S_w$ where $w$ is the total number of observation windows.
\end{enumerate}

\emph{weighted.} Given the impossibility of necessarily observing all of the tiles as they rise and set given the requirement of using multiple exposures per tile, we are motivated to define a scheme whereby each tile is given a weight based on both gravitational-wave likelihood enclosed, the number of exposures required for that tile, and the number of available slots for it to be image. Therefore we define the weights $w_i$ as
\begin{equation}
w_i = L_\textrm{GW}(\alpha_i,\delta_i) \times \frac{N_\textrm{R}}{N_\textrm{A}} 
\end{equation}
Therefore, for each exposure segment, we calculate the weight for each possible tile and select the tile with the highest weight to fill that slot.
\subsection{Efficiency}
\begin{lstlisting}
python gwemopt_run --doEvent --doPlots --doTiles --doSchedule --doEfficiency
\end{lstlisting}
We are able to test and compare the performance of these algorithms by performing simulated observations. 
We adopt observational constraints as follows. 
We use an observing limit of an altitude of 30$^\circ$, corresponding to an airmass of 2.0. 
We assume observations are available to begin at twilight and dawn, corresponding to when the sun is 12$^\circ$ below the western and eastern horizons.
We do not point away from the moon or account for sky brightness.

To estimate the efficiency for the ``detection'' of the electromagnetic counterparts to gravitational-wave transients, we perform simulated injections of supplied lightcurves. We provide example lightcurves for a variety of lightcurve models, including:

\begin{enumerate}
\item \cite{TaHo2014}: Simulations of binary systems showing ejecta morphology and resulting lightcurves. These simulations led to analytical models for black-hole neutron star systems from \cite{KaKy2016} and\cite{DiUj2017}.
\item \cite{KaKy2016}: Analytical models for black-hole neutron star systems based on \cite{TaHo2014}
\item \cite{DiUj2017}: Analytical models for binary neutron star systems based on \cite{TaHo2014}
\item \cite{BaKa2016}: Simulations of binary systems studying the emission profiles of radioactive decay products from the merger.
\item \cite{MeBa2015}: Blue ``precursor'' to the kilonovae driven by $\beta$-decay of the ejecta mass.
\item \cite{Me2017}: toy model with grey opacity for lanthanide-free matter with a density profile expanding with a range of velocities with $M(< v) = v^{-1}$.
\end{enumerate}

The requirements for ``detection'' of the electromagnetic counterparts to gravitational-wave transients are as follows.
We require that the transient appear in 2 images over 2 nights.
In each image, the transient must exceed the limiting magnitude in that image.
The color of the transient is estimated from the filter given in the configuration file.
We simulate the transients at a variety of location and distances consistent with the gravitational-wave probability skymap.

\section{Performance}
\label{sec:performance}
In this section, we compare the efficiency of the algorithms based on simulated information about what percentage of the events the algorithm can detect. According to the workflow given in Figure~\ref{fig:flowchart} and the algorithms given in the sections above, we will have four options for tiling algorithms, three options for time allocation algorithms and another three options for scheduling algorithms. This combines to 36 total options for the whole workflow. We want to know which combination has the best efficiency and then analyze and compare the algorithms individually.
\subsection{Method}
\begin{figure}[t]
\centering
\includegraphics[width=0.5\textwidth]{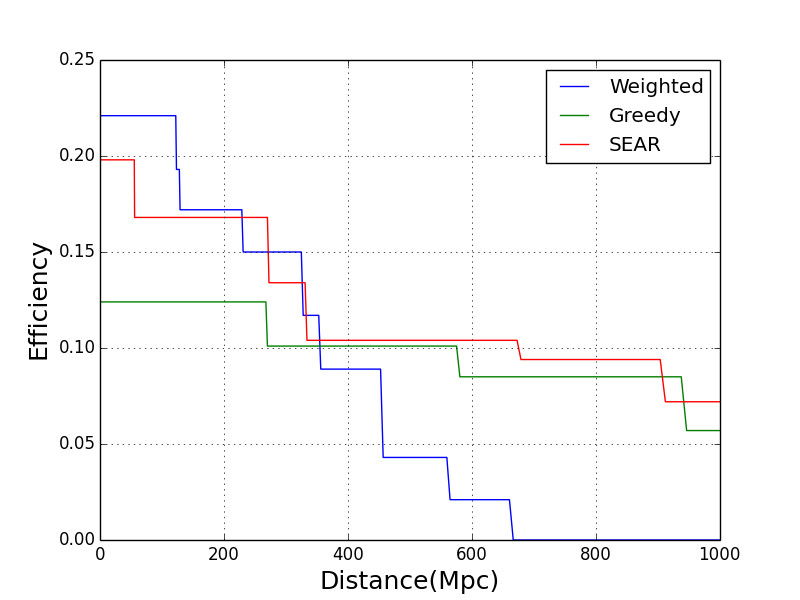}
\caption{Example plot of efficiency for \cite{Me2017} injections, comparing the scheduling algorithms. Greedy algorithm is used for tiling and PEM algorithm is used for time allocation. Greedy and SEAR have similar performance in long distance and both are better than weighted. This difference is also reflected in Figure~\ref{fig:eff_metric}. As the algorithm accounts for observability from a site, including both whether tiles are visible from the site of interest as well as diurnal effects, efficiencies are expected to peak at around 25\% for an event which fades quickly and has a probability region with peaks in both the north and south.}
\label{fig:eff_ex}
\end{figure}

We will focus on the model by \cite{Me2017} to compare the efficiency. All the efficiency values in the distance range between $10^{-1} Mpc$ to $10^3 Mpc$ (logarithmically spaced) are calculated and plotted. Thus we will have a plot of efficiency against distance for each of the 36 algorithm combinations. An example efficiency plot of efficiency is shown in Figure~\ref{fig:eff_ex}, where difference time allocation algorithms are compared. Greedy algorithm is used for tiling and PEM algorithm is used for time allocation. It can be seen that greedy and sear scheduling does better than weighted at long distance. 

In order to compare the 36 efficiencies as plotted in Figure~\ref{fig:eff_ex}, we use a single statistic to reflect the overall performance of the algorithms based on the efficiency for each distance in the range of $10^{-1} Mpc$ to $10^3 Mpc$. Thus, we come to a metric that reflects what percentage of events that can be detected in a spherical volume of radius $10^3 Mpc$. The events are evenly distributed in the volume. Suppose the event density per volume is $\rho$ and the distance is $r$. Sampling a distance at $d$ corresponds to a shell with volume $4 \pi r^2 dr$. Assuming that the density is $\rho$, then the total events on that shell is $4 \pi \rho r^2 dr$. Thus, if the efficiency is $e$, the expected number of detected events on the shell will be $4 \pi \rho r^2 e dr$. From this we can see that the efficiency at $r$ is weighted by $r^2$. If we treat the efficiency at each distance as an individual sample, a weighted average on the squared radius will then be a good metric of the overall efficiency that reflects how well the algorithm detects events uniformly distributed in a volume of $10^3 Mpc$. A consequence of this metric is that the weight of $r^2$ makes the long range efficiency more important than short range efficiency. Under this metric, an algorithm that performs well at further distances would be better than an algorithm does better at short distances but whose performance deteriorates quickly as the distance increases. 

\subsection{Performance and algorithms}
\begin{figure}[t]
\includegraphics[width=0.45\textwidth]{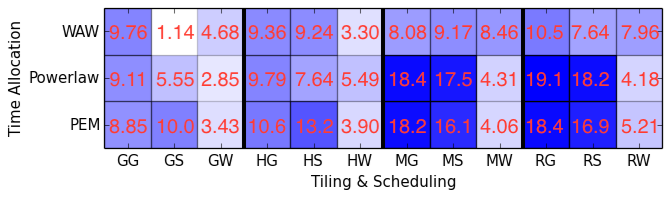}
\caption{Plot of the efficiency metric for each of the 36 options. On the horizontal axis are tiling algorithms and scheduling algorithms and on the vertical axis are the time allocation algorithms. Abbreviations are used for the algorithms. The first capital letter stands for the tiling algorithm and the second letter stands for the scheduling algorithm. The abbreviations are the first letters of the algorithms: G - greedy. H - hierarchical. M - MOC. R - ranked. S - SEAR. W - weighted. The grids are colors such that highest efficiency combinations are darker and lower efficiency ones are lighter, with the highest being completely blue and the lowest one being completely white.}
\label{fig:eff_metric}
\centering
\end{figure}
For each of the 36 algorithm options, we compute the efficiency metric as described above, which results in 36 numerical efficiency values. The results are plotted in Figure~\ref{fig:eff_metric}. On the horizontal axis are the combined options for the tiling algorithm and scheduling algorithm. There are four tiling algorithms and three scheduling algorithm so they combine to 12 columns on the horizontal axis. Abbreviations are used for the algorithms. The first capital letter stands for the tiling algorithm and the second letter stands for the scheduling algorithm. On the vertical axis are the time allocation algorithms. The color in the 36 boxes shows the efficiency as measured above. The colors are scaled to the efficiency such that higher efficiencies are more darkly colored.
The highest efficiency of $0.19$ is achieved by a combination of ranked tiling, powerlaw time allocation and greedy scheduling. Compared to the lowest efficiency of $0.01$, it can detect roughly 19 times more events within a range of $10^{-1} Mpc$ and $10^3 Mpc$. That corresponds to the darkest box in the 10th column and the second row in Figure~\ref{fig:eff_metric}. 
Also, from Figure~\ref{fig:eff_metric}, we can compare the efficiencies of the individual algorithms. First, among the four tiling algorithms, greedy, sear, MOC and ranked, we can see that MOC and ranked generally have higher efficiencies than greedy and hierarchical tiling algorithms. Second, among the three options for scheduling, greedy and SEAR have higher efficiency. Generally, we can say that PEM gives the best results. It is slightly better than powerlaw, and both are better than WAW. This is unsurprising as PEM is optimal in the presence of distance information, while WAW requires inclination information in order to be optimal. However, in case inclination and distance were available in low latency, WAW may be best. Third, greedy and SEAR scheduling are more efficient than weighted scheduling.
\subsection{Performance and the number of tiles}
We also study the how the number of tiles affects the performance. It is only relevant in the hierarchical and greedy tiling algorithm. The hierarchical algorithm is used for this study, which selects tiles that covers the highest probability and mask the tiles once they are selected. It stops when a user-defined number of tiles is selected. The efficiency is computed based on the simulation of 1000 injections.

\begin{figure*}[t]
    \centering
    \includegraphics[width=0.45\textwidth]{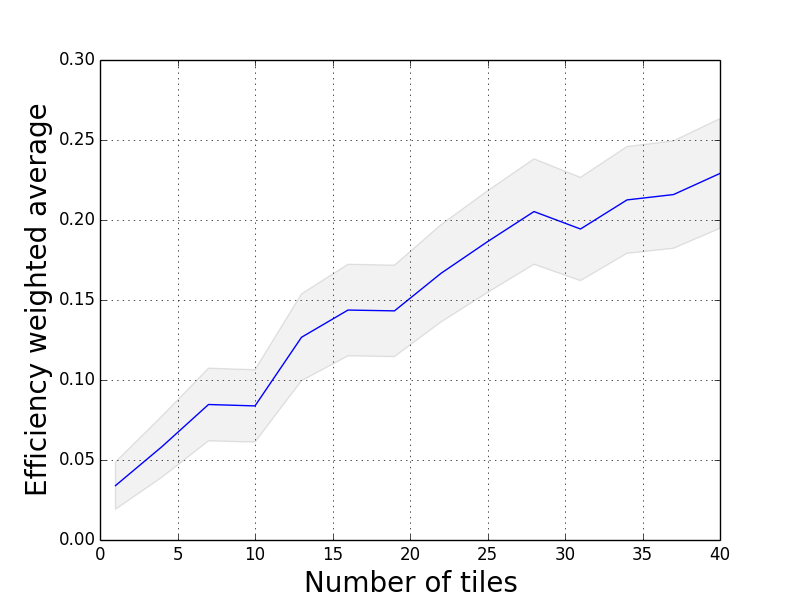}
    \includegraphics[width=0.45\textwidth]{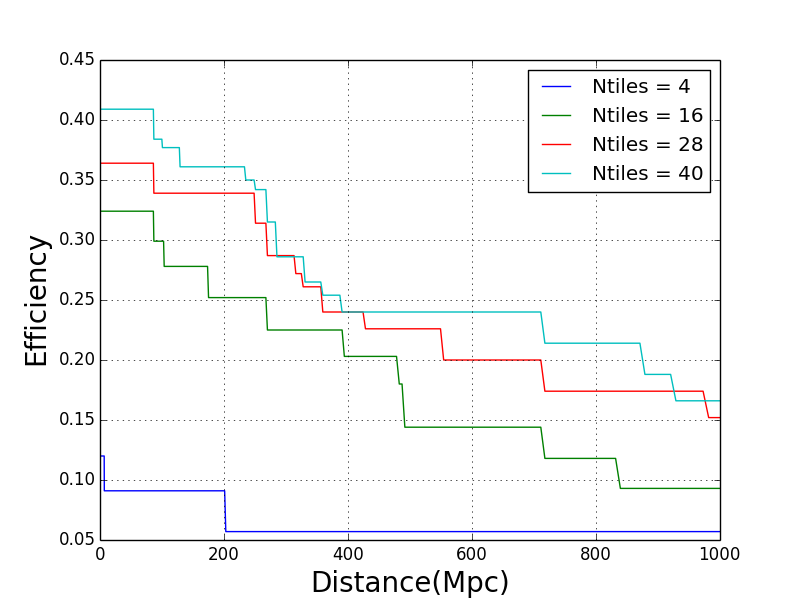}
    \caption{On the left is the plot of efficiency weighted average as a function of the number of tiles considered. On the right is the efficiency curves unpacked from data points corresponding to 4, 16, 28, 40 tiles of the top figure. In the simulation, hierarchical tiling algorithm, power law time allocation algorithm and greedy scheduling algorithm are used to detect 1000 random injections for number of tiles ranging from 1 to 40. The standard deviation is calculated and the 99\% confidence interval is plotted as the grey shaded region in the top figure.}
    \label{fig:eff_ntiles}
\end{figure*}

From the plot on the left of Figure~\ref{fig:eff_ntiles}, it can be seen that considering more tiles generally results in better efficiency. However, the benefit of increasing the number of tiles decreases as more and more tiles are considered. This effect is shown in the left plot, where the efficiency curves are shown for different numbers of tiles. The three lines showing the results of 16, 28 and 40 tiles are closer to each other than the bottom one where only four tiles are selected. This means that increasing the number of tiles from 4 to 16 improves the efficiency more than from 16 to 28. This is expected since, as the number of tiles increases, new tiles cover less and less probability and the limiting factors in the time allocation and scheduling algorithm become more important and the efficiency will not keep rising.
Note that the number of tiles in our study is the direct output of the tiling algorithm. A tile still needs to go through the time allocation and scheduling algorithm to be scheduled for observation. It is independent of the CCD readout time and telescope slew. For an alternative study that accounts for these effects, please see Chan et al. \cite{ChHu2017}.

\section{Conclusion}
\label{sec:conclusions}

The detection of GW170817 \citep{AbEA2017b} has invigorated the search for improved strategies for associating gravitational waves with electromagnetic
counterparts.
Due to the large uncertainty footprint, which can range from 100-1000 square degrees, efficiently scanning sky areas  of this size in search of an electromagnetic counterpart is challenging.
However, we have described in this paper a number of algorithms in the literature available for significantly improving upon the most naive approach.
We have shown comparisons between the algorithms, describing the limits in which they are the most effective.

One potential improvement to the analysis considered here is using the locations of known galaxies in the gravitational-wave sensitivity volume, which was $\approx 100$\,Mpc for GW170817 \citep{AbEA2017b} and will extend to $\approx 300$\,Mpc at design sensitivity \citep{aLIGO}. 
Recent improvements in galaxy catalog completeness have made this effort possible. For example, the Galaxy List for the Advanced Detector Era (GLADE) galaxy catalog is complete (with respect to a Schechter function) out to $\approx 300$\,Mpc for galaxies brighter than the median Schechter function galaxy luminosity \footnote{http://aquarius.elte.hu/glade/}. The Census of the Local Universe (CLU) catalog \citep{CoKa2017} is complete to 85\% in star-formation and 70\% in stellar mass at 200\,Mpc.
Within these local volumes, the sky area coverage of galaxies is $\approx 1$\,\% \cite{CoKa2017}, bringing the sky areas searched down by a factor of 100, which makes the possibility of targeted galaxy pointing tractable, especially for small field of view telescopes (see \cite{ArMc2017} for an example). 

A code to produce the results in this paper is available at 
\url{https://github.com/mcoughlin/gwemopt} for public download.

\section{Acknowledgments}
MC is supported by the David and Ellen Lee Prize Postdoctoral Fellowship at the California Institute of Technology. 
DT and NC are supported by the National Science Foundation, under NSF grant number PHY 1505373.
YH is supported by the National Natural Science Foundation of China, under NSFC 11703098.
CWS is grateful to the DOE Office of Science for their support under award DESC 0007881.

\bibliographystyle{aasjournal}
\bibliography{references}

\end{document}